\documentclass[prl,aps,twocolumn,10pt,showpacs]{revtex4-1}
\usepackage{graphicx}
\usepackage{amsmath}

\begin{document}

\title{Reply to ``Comment on `A Scaling law beyond \\ Zipf's law and its relation to Heaps' law' ''}

\author{Francesc Font-Clos$^{1,2}$}
\author{\'Alvaro Corral$^{1,2}$}

\affiliation{
$^{1}$Centre de Recerca Matem\`atica,
Edifici C, Campus Bellaterra,
E-08193 Barcelona, Spain.\\
$^{2}$Departament de Matem\`atiques,
Facultat de Ci\`encies,
Universitat Aut\`onoma de Barcelona,
E-08193 Barcelona, Spain
}

\maketitle

In Ref. \cite{Minnhagen2009}, Bernhardsson, da Rocha, and Minnhagen
proposed that the distribution of word 
frequencies in a text or collection of texts (of the same author) changes
with text length as
\begin{equation}
D_L(k) = A \frac{e^{-k/(c_0 L)}}{k^{\gamma(L)}},
\label{eq:scalingsuecos}
\end{equation}
where 
$k$ is the absolute frequency (number of tokens)
of the different words (word types),
$L$ is text length in number of tokens ($M$ in Ref. \cite{Minnhagen2009}'s notation),
$D_L(k)$ is the probability mass function of $k$
(i.e., the distribution of word frequencies), 
$\gamma(L)$ is the power-law exponent, 
$c_0$ is a scale parameter (independent on $L$),
and $A$ is a normalizing constant.
The key ingredient of Bernhardsson et al.'s approach
to model the change of $D_L(k)$ with $L$
is the explicit dependence of the exponent $\gamma$
on text length $L$, decreasing with increasing $L$.

Alternatively, in Ref. \cite{Font-Clos2013}, we argue that 
the variability of the statistics of words in a text with its
size is better explained by a scaling law,
\begin{equation}
\label{eq:scalingfontclos}
D_L(k) = \frac 1 {L V_L} g(k/L),
\end{equation}
where $V_L$ is the size of vocabulary
(number of different words, i.e.,  word types),
and $g(x)$ is a undefined scaling function, 
independent of text size.

Now, Yan and Minnhagen \cite{Yan_comment} claim that our scaling law is 
``fundamentally impossible'' and ``fundamentally incorrect''.
We summarize the points of these authors to make it clear
that their criticism is essentially irrelevant.
First, in Fig. 1, they find that our scaling law does not hold 
for $k=1$.
Second, in Fig. 2 they show that our scaling does not work well for,
let us say, $k \le 10$.
Third, it is argued that a ``Randomness view'', based
in the concepts of 
``Random Group Formation'', ``Random Book Transformation'', and 
``Metabook'' predicts the right form of $D_L(k)$,
which is that of Ref.\cite{Baek2011}.

It is obvious that the first and second criticisms of Yan and Minnhagen
are irrelevant, as they simply imply that our scaling law can only be
valid beyond the low-frequency limit, so, 
$$
D_L(k) = \frac 1 {L V_L} g(k/L), \mbox{ for } k > 10.
$$
This is not surprising at all, as it is well known in statistical physics
that scaling laws hold asymptotically.
It is remarkable that, for texts, scaling is attained after the first
decade in frequencies.
It is also remarkable that, despite the fact that Yan and Minnhagen 
stretch the scaling hypothesis up to very small fragments of texts
($212473/500 \simeq 400$ tokens, for the case of \emph{Moby-Dick}), the scaling law still is fulfilled reasonably well, 
beyond the first decade in $k$.
Naturally, the appropriate way to further test the validity of our scaling
law is in the opposite way, analyzing larger and larger texts.

\begin{figure*}
\begin{center}
\includegraphics[width=\textwidth]{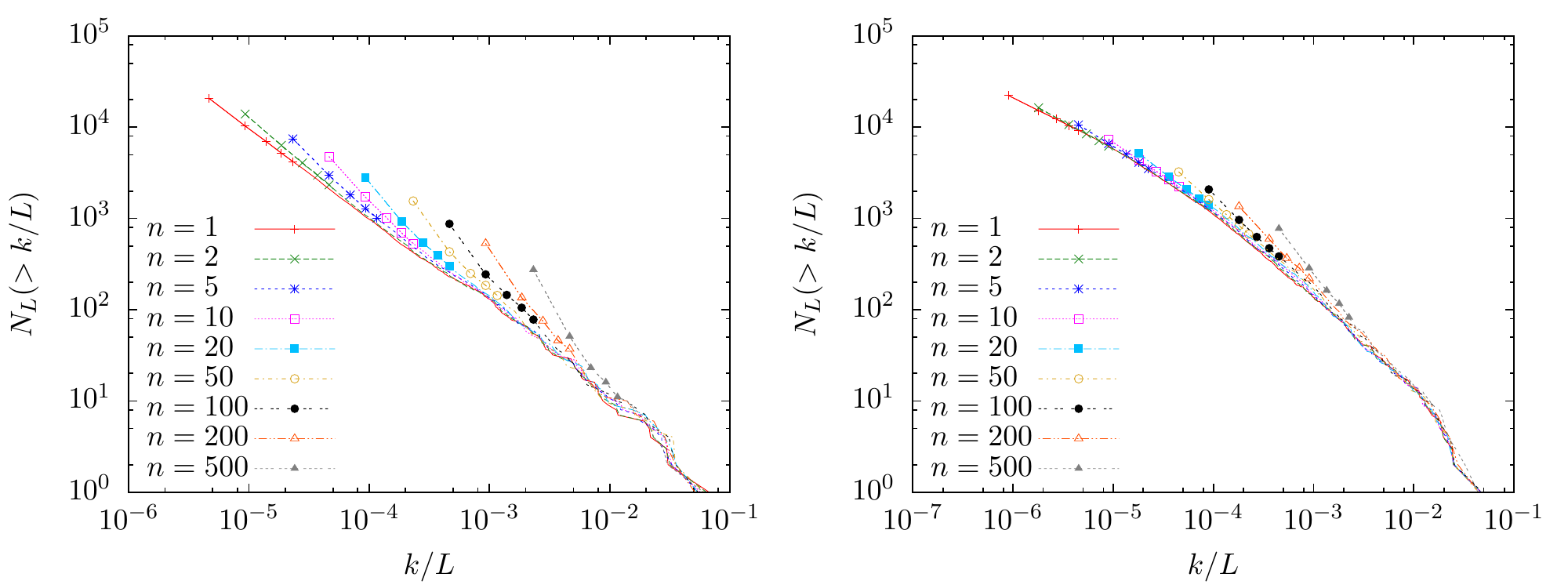}
\end{center}
\caption{
\label{fig:5points}
The total number of words $N_L$ with a relative frequency greater than or equal to $k/L$, for varying $L=L_{\textrm{tot}}/n$. We have taken the same books as in Ref. \cite{Yan_comment}, \emph{Moby-Dick} (left) and \emph{Harry Potter} (right), exactly reproducing panels (a) and (b) of Fig. 2 in Ref. \cite{Yan_comment}, but also including some additional values of $n$. Lines are drawn for all $k$, but symbols are drawn only for $k=1\dots 5$, showing that deviations from the scaling law are always in the regime of very low frequencies, as expected due to discreteness effects. 
}
\end{figure*}

\begin{figure*}
\begin{center}
\includegraphics[width=\textwidth]{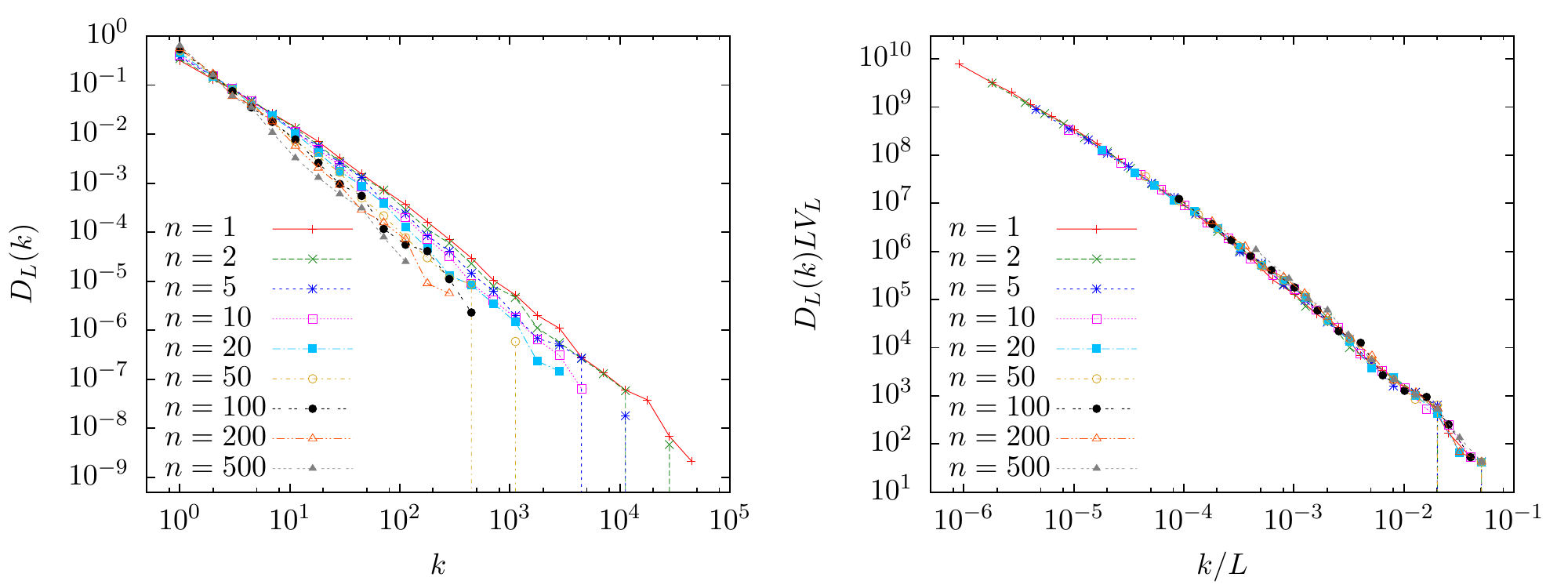}
\end{center}
\caption{
\label{fig:harrypotter}
\textbf{Left:} The probability mass function $D_L(k)$ of the absolute frequency $k$, for varying subsets of length $L=L_{\textrm{tot}}/n$ of \emph{Harry Potter}, displaying a seeming change of shape. \textbf{Right:} Same, but plotting $D_L(k) L V_L$ versus $k/L$, as proposed in Ref. \cite{Font-Clos2013} and stated here in Eq.~\eqref{eq:scalingfontclos}. All curves collapse into a single, length-independent scaling function $g(k/L)$, in agreement with Eq.~\eqref{eq:scalingfontclos}, and at odds with Eq.~\eqref{eq:scalingsuecos}: a length-dependent exponent in  $D_L(k)$, as proposed by Yan and Minnhagen, is not compatible with the data collapse shown in the figure.
}
\end{figure*}

To make our point more clear, in Fig. \ref{fig:5points} we  present the same data as in Fig. 2 of Ref. \cite{Yan_comment}, but adding symbols for $k=1\dots5$ (instead of only lines, as in Ref. \cite{Yan_comment}). It is apparent that even in the extreme case of $n=500$, the scaling law only fails for very small frequencies. Additionally, in Fig. \ref{fig:harrypotter} we perform the data collapse associated to our scaling law for the case of \emph{Harry Potter}, presented in Ref. \cite{Yan_comment} as a counter-example to our scaling law. As it is shown, the collapse is excellent: after proper rescaling, all curves collapse into a single, length-independent function, even for very small frequencies.

So, the empirical facts are clear: a scaling law gives a very good approximation for the distribution of word frequencies in the range
$k > 10$. If the ``Randomness view'' hold by Yan and Minnhagen 
is valid, then it must contain in some limit the scaling law.
If not, their theory is wrong. As a final remark, let us state that
although curve fitting is a very honorable approach in science
(when done correctly \cite{Corral_Deluca}), our scaling approach has nothing to do with that,
contrary to Yan and Minnhagen's claims. 

In summary, the objections raised by Yan and Minnhagen 
are too weak to justify the publication of a comment to our work.

\addcontentsline{toc}{chapter}{Bibliography}
\vspace*{-3mm}
\bibliography{biblio}
\bibliographystyle{unsrt}

\end{document}